\documentclass[english,english,aps,prd,amsmath,amssymb,superscriptaddress]{revtex4}
\usepackage[T1]{fontenc}
\usepackage[latin9]{inputenc}
\usepackage{amsmath}
\usepackage{amssymb}
\usepackage{esint}

\makeatletter
\@ifundefined{textcolor}{}
{%
 \definecolor{BLACK}{gray}{0}
 \definecolor{WHITE}{gray}{1}
 \definecolor{RED}{rgb}{1,0,0}
 \definecolor{GREEN}{rgb}{0,1,0}
 \definecolor{BLUE}{rgb}{0,0,1}
 \definecolor{CYAN}{cmyk}{1,0,0,0}
 \definecolor{MAGENTA}{cmyk}{0,1,0,0}
 \definecolor{YELLOW}{cmyk}{0,0,1,0}
}

\usepackage{slashed}
\usepackage{babel}
\usepackage{bbm}

\makeatother

\usepackage{babel}
\begin{document}

\title{Quark-Gluon Plasma and Topological Quantum Fields Theory}

\author{M.J.Luo}

\email{mjluo@ujs.edu.cn}

\affiliation{Department of Physics, Jiangsu University, Zhenjiang, Jiangsu 212013,
China}
\begin{abstract}
Based on an analogy with topologically ordered new state of matter
in condensed matter systems, we propose a low energy effective field
theory for a parity conserving liquid-like quark-gluon plasma (QGP)
around critical temperature in quantum chromodynamics (QCD) system.
It shows that below a QCD gap which is expected several times of the
critical temperature, the QGP behaves like topological fluid. Many
exotic phenomenon of QGP near the critical temperature discovered
at RHIC are more readily understood by the suggestion that QGP is
a topologically ordered state.
\end{abstract}
\maketitle

\section{Introduction}

Quantum Chromodynamics (QCD) based on non-Abelian gauge field theory
had been well established as the fundamental theory of the strong
interaction. However, because of its strong coupling nature at low
energy or long distance, from the day the exact theory was written
down, its low energy behavior is still being poorly understood by
solving the first principle theory. Although we already have a low
energy effective theory based on the Goldstone's theorem under the
condition of chiral symmetry breaking (for a review, see: \cite{scherer2003introduction}),
there is no analog foundation to formulate a low energy effective
theory with neither symmetry breaking nor order parameter, more specifically,
applying to the disordered quark-gluon plasma (QGP) phase. 

A crucial phenomenological clue to the problem comes from the experimental
studies of the Relativistic Heavy Ion Collision (RHIC), in which new
state of matter of strong interaction is produced. It is discovered,
at sufficiently high temperature and density, that the nucleons in
the collision are deconfined into constituent quarks and gluons and
form a soup of QGP. Important experimental facts of the QGP are the
following: (i) The long distance low frequency behavior of the QGP
around the critical temperature $T_{c}$ is well described by the
theory of fluid mechanics (hydrodynamics) \cite{kolb2001elliptic,kolb2003transverse}.
(ii) The QGP around $T_{c}$ looks like a strong correlated and near-perfect
liquid \cite{shuryak2005rhic}, the pressure, numbers of transport
properties (e.g. conductivity, diffusion and viscosity) are small,
strongly deviating from the behavior of a weakly coupled ideal gas
predicted from the perturbative QCD. (iii) The charge of the deconfined
quark or parton excitation in the QGP is fractionalized \cite{jeon2000charged}.
In the temperature range $T_{c}<T\lesssim2T_{c}$ where these experimental
facts are established, we propose a topological dominant effective
field theory for the parity conserving QGP by bosonizing the gauge
invariant QCD that I feel are consistent with the known experimental
facts.

The connection between a pure gluon plasma in hard thermal loop limit
and the topological Chern-Simons theory has been discussed in literature
\cite{efraty1992action,efraty1993chern}. The remarkable relation
between the QGP and topological field theories can be generalized
to a more generic level. The QGP phase is a deconfined phase where
the bound state hadrons are decomposed into quarks and gluons at high
temperature and density. The situation is analogous with the case
in condensed matter physics such as the gauge symmetric Heisenberg
spin system that at high enough temperature and chemical potential
where the system is well into a deconfined phase, the Heisenberg spins
are decomposed into constituent holons and spinons and form a plasma
phase which enjoy a new type of order called topological order/phase
(see, for example, \cite{wen1991mean,wen2002quantum} at low dimensions,
\cite{diamantini2011charge,vishwanath2013physics,chan2013effective}
in 3+1 dimensions). The ground state of the topologically ordered
matter is not associated with spontaneous symmetry breaking, protected
by a gap, and topological fluid (see, for example, \cite{frohlich1991large,zee1992long,wen1992classification}
in 2+1 dimensions) as the low energy degree of freedom in many aspects
is analogous with the fluid-like plasma of quark-gluon around the
critical temperature and chemical potential. The topological fluid
is in general incompressible, with excitation carrying fractional
charges \cite{laughlin1983anomalous,haldane1983fractional,jain1989incompressible},
characterized by having non-trivial ground state degeneracies \cite{wen1990ground,wen1991mean}
on topological non-trivial compact manifold. 

Bosonization is a proper description to such bosonic fluids system,
especially at finite chemical potential because of a more controllable
sign structure than fermion \cite{wu2008sign,luo2014gauge}. The validity
of bosonization approach in the real 3+1 dimensional QCD is based
on the facts: (a) lattice data show that quark (baryon) masses seem
to be large in the strongly coupled QGP near $T_{c}$ \cite{Karsch:2009tp},
(b) a massive fermionic theory and the bosonized descriptions share
the same macroscopic limit which is dominated by topological degrees
of freedom \cite{chan2013effective}.

The structure of the paper is as follows. In section II, we construct
a basic formulation of a bosonized low energy effective field action
for the QGP. In subsection A of the section III, we derive the classical
equations of motion and argue the dyonic nature of the fluids. In
subsection B, we derive the phenomenological hydrodynamics equations
from the effective field theory. In subsection C, linear responses
of the theory are studied, more specifically, the ratio of shear viscosity
over entropy density of an homogeneous and isotropic QGP is derived
at Bogomol\textquoteright nyi bound. In subsection D, the fractionalization
of charge for quark/parton in the plasma is explained. In section
IV, we summarize the results.

\section{Bosonic Effective Action}

Both for mathematical and physical reasons, in a strongly coupled
many-body system, what we observed are gauge invariant responses to
external (in practice electromagnetic) probe, and hence the notion
of response, especially linear response, is more proper than the unobserved
microscopic fundamental particles. The bosonization approach starts
from replacing the fermionic fundamental quarks by the responses to
an external weak U(1) gauge field being a source minimally coupling
to them,
\begin{equation}
\left\langle j_{\mu_{1}}(x_{1})j_{\mu_{2}}(x_{2})...j_{\mu_{n}}(x_{n})\right\rangle =\frac{\delta}{\delta A_{\mu_{1}}^{ex}(x_{1})}\frac{\delta}{\delta A_{\mu_{2}}^{ex}(x_{2})}...\frac{\delta}{\delta A_{\mu_{n}}^{ex}(x_{n})}\ln Z[A^{ex}],\label{eq:current-correlation}
\end{equation}
where
\begin{equation}
Z[A^{ex}]=\int\mathcal{D}\bar{\psi}\mathcal{D}\psi\mathcal{D}G\exp\left(-\int d^{4}x\bar{\psi}\left[i\gamma_{\mu}\left(\partial_{\mu}+iG_{\mu}+\mu\delta_{\mu0}+iA_{\mu}^{ex}\right)\right]\psi-S_{YM}[G]\right).
\end{equation}
The $\psi$ are quarks, $G_{\mu}$ are gluon gauge fields, $\mu$
is the chemical potential, $S_{YM}[G]$ is the Yang-Mills action for
gluons, and $A_{\mu}^{ex}$ are external U(1) source coupled to an
Abelian subgroup of the Non-Abelian $SU(N_{c})$ of QCD.

The form of $Z[A^{ex}]$ can be determined as follows. By using the
U(1) gauge invariance $Z[A^{ex}]=Z[A^{ex}+a]$. $a$ is a to-be-interpreted
(see following) 1-form gauge field in the effective QCD under the
Abelian projection of $SU(N_{c})\sim U(1)^{N_{c}-1}$ which fixes
the gauge partially. So there are $N_{c}-1$ such fields, and here
$a$ describes an effective mixing of those coupled to $A^{ex}$.
And we consider it is a pure gauge, i.e. $f_{\alpha\beta}(a)=\partial_{\alpha}a_{\beta}-\partial_{\beta}a_{\alpha}=0$.
The pure gauge condition can be imposed by inserting a delta function
into the functional integral,
\begin{align}
Z[A^{ex}] & =\int\mathcal{D}aZ[A^{ex}+a]\left(\prod_{x}\prod_{\mu<\nu}\epsilon_{\mu\nu\alpha\beta}\delta[f_{\alpha\beta}(a(x))]\right).
\end{align}
The delta function can be exponentiated by introducing an auxiliary
anti-symmetric 2-form gauge fields $b_{\mu\nu}$ as Lagrangian multipliers,

\begin{equation}
Z[A^{ex}]=\int\mathcal{D}a\mathcal{D}bZ[A^{ex}+a]\exp\left(-\frac{i}{4\pi}\int d^{4}x\epsilon_{\mu\nu\alpha\beta}b_{\mu\nu}f_{\alpha\beta}(a)\right),
\end{equation}
in which the term in the exponent is purely topological known as the
BF action (for a review see: \cite{birmingham1991topological}), and
the pre-factor $\frac{1}{4\pi}$ is for convention. The pre-factor
can be in general defined as $\frac{\mathsf{m}}{4\pi}$, $\mathsf{m}\in\mathbb{Z}$
plays a similar role of a Chern-number in quantum Hall systems characterizing
its topological order, the integer-valued $\mathsf{m}$ so defined
characterizes the topological order of the ground state of the system,
a value $\mathsf{m}\neq1$ represents a fractionalized topological
order. In such setting, under a gauge transformation the BF term is
just shifted by an integer multiple of $2\pi$, so the partition function
is not affected.

The auxiliary gauge field $b$ is of ($D-2$)-form where $D$ is spacetime
dimensions. In the paper we will focus on the real QCD system $D=4$,
so $b$ field is of a 2-form. By doing a variable replacement $a\rightarrow a-A^{ex}$,
we obtain
\begin{equation}
Z[A^{ex}]=\int\mathcal{D}a\mathcal{D}bZ[a]\exp\left(-\frac{i}{4\pi}\int d^{4}x\epsilon_{\mu\nu\alpha\beta}b_{\mu\nu}\left[f_{\alpha\beta}(a)-f_{\alpha\beta}(A^{ex})\right]\right).\label{eq:identity}
\end{equation}
By using Eq.(\ref{eq:current-correlation}), the above partition function
indicates a bosonized flow
\begin{equation}
j_{\mu}(x)=\frac{1}{2\pi}\epsilon_{\mu\nu\alpha\beta}\partial_{\nu}b_{\alpha\beta}(x),\label{eq:charge-current}
\end{equation}
which is gauge invariant under the transformations
\begin{equation}
b_{\mu\nu}\rightarrow b_{\mu\nu}+\partial_{\mu}\eta_{\nu}-\partial_{\nu}\eta_{\mu}.
\end{equation}
Note that the flow is automatically conserved 
\begin{equation}
\partial_{\mu}j_{\mu}\equiv0.\label{eq:charge-current-conservation}
\end{equation}
The non-dynamical component $a_{0}$ coupled to the density $j_{0}$
plays the role of a Lagrangian multiplier similar with a chemical
potential.

Beside $a_{0}$, there is another non-dynamical Lagrangian multiplier
$b_{0i}=-b_{i0}$, ($i=1,2,3$), which implies the existence of a
spin/angular momentum or magnetization density tensor 
\begin{equation}
s_{\mu\nu}(x)=\frac{1}{2\pi}\epsilon_{\mu\nu\alpha\beta}\partial_{\alpha}a_{\beta}(x).\label{eq:spin-density}
\end{equation}
It can be interpreted as a spin/angular momentum or magnetization
density because it is a response to an external magnetic field which
will be shown as following. If we add a kinetic term of $b_{\mu\nu}$
with infinite mass into the parenthesis of the Eq.(\ref{eq:identity})
as a regulator,
\begin{equation}
S\rightarrow S+S_{reg},\quad S_{reg}=\lim_{\lambda\rightarrow\infty}\frac{1}{12\lambda^{2}}\int d^{4}xh_{\mu\nu\rho}h_{\mu\nu\rho},\label{eq:kinetic-b}
\end{equation}
where $h_{\mu\nu\rho}=\partial_{\mu}b_{\nu\rho}+\partial_{\nu}b_{\rho\mu}+\partial_{\rho}b_{\mu\nu}$
is the field strength of $b_{\mu\nu}$, and $\lambda$ is a mass parameter.
By integrating out the $b_{\mu\nu}$ field, we obtain an effective
coupling between $s_{\mu\nu}$ and external source field, 
\begin{equation}
S_{int}=\frac{1}{4\pi}\int d^{4}xs_{\mu\nu}\tilde{F}_{\mu\nu}^{ex},
\end{equation}
in which $\tilde{F}_{\mu\nu}^{ex}=\epsilon_{\mu\nu\alpha\beta}\partial_{\alpha}A_{\beta}^{ex}$
is an external magnetic field. So we prove that the tensor $s_{\mu\nu}$
is a response to the external magnetic field,
\begin{equation}
\left\langle s_{\mu_{1}\nu_{1}}(x_{1})s_{\mu_{2}\nu_{2}}(x_{2})...s_{\mu_{n}\nu_{n}}(x_{n})\right\rangle =(4\pi)^{n}\frac{\delta}{\delta\tilde{F}_{\mu_{1}\nu_{1}}^{ex}(x_{1})}\frac{\delta}{\delta\tilde{F}_{\mu_{2}\nu_{2}}^{ex}(x_{2})}...\frac{\delta}{\delta\tilde{F}_{\mu_{n}\nu_{n}}^{ex}(x_{n})}\ln Z[A^{ex}].\label{eq:spin-correlation}
\end{equation}
In the rest of the paper we will not strictly distinguish the terminologies
spin or angular momentum or magnetization, but using ``spin'' for
convention, and a corresponding currents of $s_{\mu\nu}$ we call
them spin currents. 

The spin currents can be defined by a spin density tensor
\begin{equation}
M_{[\mu\nu]\alpha}=\partial_{\alpha}s_{\mu\nu}.\label{eq:spin-current}
\end{equation}
From its definition Eq.(\ref{eq:spin-density}) and Eq.(\ref{eq:spin-current}),
we can see that the spin currents satisfies 6 continuity equations
\begin{equation}
\partial_{\alpha}M_{[\mu\nu]\alpha}\equiv0.\label{eq:spin-current-conservation}
\end{equation}
These continuity equations ensure the spin currents are conserved
or/and divergentless. In addition, there is another continuity equation,
note that a Pauli-Lubanski pseudovector constructed by the spin current
\begin{align}
W_{\mu}(x) & =\frac{1}{2}\epsilon_{\mu\nu\alpha\beta}M_{[\alpha\beta]\nu},
\end{align}
is also automatically satisfied 
\begin{equation}
\partial_{\mu}W_{\mu}\equiv0.\label{eq:Pauli-conservation}
\end{equation}

The independence of the continuity equations for charge and spin currents
manifests an important fact that the carriers of charge and spin in
the fluids are no longer identified, in contrast to the conventional
quasi-particle picture in which it carries both. In other words, there
are two types of currents in the fluids \cite{diamantini2011charge},
which is known as spin-charge separation in condensed matter physics.

So far the bosonization recipe is quite general and is able to apply
to a gapped or gapless system, but only in 1+1 dimensions the $Z[a]$
can be written down exactly, here for 3+1 dimensions $Z[a]$ can only
be approximately evaluated below a cutoff or gap $\Lambda$. The $Z[a]$
now encodes almost all physics of QCD, which needs some guess work.
The bosonization gap $\Lambda$ is closely related to the gap of the
fermionic spectrum, from the lattice date \cite{zwanziger2005equation,Karsch:2009tp}
it is shown that the gap is at least several time larger than $T_{c}$
when the temperature of sQGP is around $T_{c}$, or even much larger
\cite{Shuryak:1999qs,Zakharov:2002md}). In the realistic heavy-ion
collision experiments, the gap could be different in different situations:
e.g. $\Lambda\sim\sqrt{T^{2}+\mu^{2}/\pi^{2}}$ at high temperature
and chemical potential $\mu$, $\Lambda\sim\sqrt{e\left|B\right|}$
at strong magnetic fields. In short, it is safe to consider that $\Lambda$
will be larger than the temperature range of the validity for our
low energy effective field theory $T_{c}<T\lesssim2T_{c}$ for a typical
hydro-like sQGP at RHIC, so that the phenomenological results predicted
well below the gap will not depend on the exact value of $\Lambda$.
Therefore we could expand $Z[a]$ by the inverse of the gap, two most
general leading terms constructed by the gauge fields $a$ can be
explicitly written down based on the principle of gauge invariance,
i.e. a $\theta$-term and a Maxwell term. The two leading terms have
no dimensional parameter in 3+1 dimensions, so they are marginal in
the RG sense. The higher order terms contain higher derivatives which
are suppressed by the gap $\Lambda$. We have
\begin{equation}
Z[a]=\exp\left(\frac{i\theta}{16\pi^{2}}\int d^{4}x\epsilon_{\mu\nu\alpha\beta}f_{\mu\nu}(a)f_{\alpha\beta}(a)-\frac{1}{4g^{2}}\int d^{4}xf_{\mu\nu}(a)f^{\mu\nu}(a)+\mathcal{O}\left(\partial^{4}/\Lambda^{2}\right)\right).\label{eq:Z=00005Ba=00005D}
\end{equation}

The first term is the $\theta$-term which is topological and always
marginal at low energy. Here the $U(1)$ symmetry for the $a$ field
is an Abelianized compact subgroup of the full $SU(N_{c})\sim U(1)^{N_{c}-1}$,
but the theory is not simply a $\mathrm{QED}^{N_{c}-1}$, something
of the non-Abelian character has to survive in the exact $Z[a]$.
As the starting point of the effective theory of QGP, it does not
associate with any symmetry breaking of QCD, we expect that the exact
$Z[a]$ encoding physics of QCD keeps invariant under $SU(N_{c})$,
and in 3+1 dimensions the homotopy $\pi_{3}[SU(N_{c})]$ is non-trivial,
so the parameter $\theta$ being an angle (with periodicity $2\pi$)
could have non-trivial choice in the theory and be physical observable.
The $\theta$-term generally breaks the parity and time-reversal symmetries
unless it is quantized as $\theta=\nu\pi$, $\nu\in\mathbb{Z}$ is
a topological non-trivial winding number. It is worth stressing that
the $\theta$ here is not necessarily the $\theta_{QCD}$ parameter
in QCD, we know $\theta_{QCD}$ is (from the dipole moment of the
neutron) exceedingly small $\theta_{QCD}<10^{-9}$. However, our following
discussions are based on a non-trivial choice 
\begin{equation}
\theta=\pi
\end{equation}
in the effective theory of QGP, which is necessary for a consistent
hydrodynamic interpretation of the topological effective fields theory.
And the choice of the value can be considered as a topological order
parameter characterizing a phase of QGP, which is an important difference
between our model and the chiral superfluid model of QGP \cite{Kalaydzhyan:2012ut,Kalaydzhyan:2014yda}.
The consequence of the non-trivial choice is an important topic in
the next section.

The second Maxwell term contains metric, so it is dynamical and non-topological.
The coupling constant $g$ in the second term effectively encodes
the information of $N_{c}$ and $N_{f}$ of the QCD, i.e. $g(N_{c},N_{f})$.
As is well known that fermions and bosons contribute opposite signs
to the $\beta$-function in renormalization: Fermions screen the coupling
constant while self-interacting bosons anti-screen it. As a result,
in the pure bosonized system there are no extra screening effects
from fermions, if the pre-bosonized theory is strongly coupled at
low energy, the bosonized version remains being strongly coupled.
For the real QCD: $N_{c}=3$, $N_{f}=6$, the theory is well-known
being asymptotic freedom or strongly coupled at low energy because
of the competition between fermions and bosons. Therefore, we expect
that the coupling constant $g$ becomes large at low energy due to
the anti-screening effects of $a_{\mu}$ field from its self-interacting.
In this case, the second term is expected small compared with the
first $\theta$-term at low energy and be considered as a perturbation,
except that a specific choice of $N_{c}$ and $N_{f}$, e.g. a Banks-Zaks
weakly coupled fixed point where $g$ is small when $N_{f}$ is close
to $11N_{c}/2$, in which case only the second term is not enough
and the contributions from higher order terms are required to be considered.
The rest paper focuses only on the strongly coupled case.

There are two remarks to $Z[a]$ taking the form as Eq.(\ref{eq:Z=00005Ba=00005D}):
(1) such expansion works at large coupling $g$ and at low energy
below the gap $\Lambda$, and the results in the rest of the paper
are achieved at strong coupling and below the gap; (2) it is such
form that is remarkably successful in re-interpreting the gauge fields
$a_{\mu}$,$b_{\mu\nu}$ as certain fluids in the ground state, and
we will see that their classical equations of motion are just hydrodynamic
equations of the fluids. In other words, such form of $Z[a]$ gives
a simple hydrodynamic re-interpretation of the QCD ground state, which
will be shown in the section-III.

So put everything together, it contains only relevant and marginal
terms and should be compatible with the gauge invariance, the bosonized
low energy effective action is given by
\begin{equation}
Z[A^{ex}]=\int\mathcal{D}a\mathcal{D}b\exp\left(-S_{eff}\right),
\end{equation}
with
\begin{equation}
S_{eff}(a,b)=\frac{i}{4\pi}\int d^{4}x\epsilon_{\mu\nu\alpha\beta}b_{\mu\nu}\left[f_{\alpha\beta}(a)-f_{\alpha\beta}(A^{ex})\right]-\frac{i\theta}{16\pi^{2}}\int d^{4}x\epsilon_{\mu\nu\alpha\beta}f_{\mu\nu}(a)f_{\alpha\beta}(a)+\frac{1}{4g^{2}}\int d^{4}xf_{\mu\nu}(a)f^{\mu\nu}(a)+\mathcal{O}\left(\partial^{4}/\Lambda^{2}\right),\label{eq:effective-action}
\end{equation}
in which the fermionic quarks are bosonized by two types of gauge
fields: 1-form $a_{\mu}$ and 2-form $b_{\mu\nu}$. The first two
terms are topological, the integrals in these terms are metric independent.
The topological terms giving rise from the bosonization play the role
of a non-trivial phase and sign structure of fermions, which is instead
responsible for the notorious sign problem for the original fermionic
system at finite chemical potential. While the third term is non-topological
and metric dependent, it is real in Euclidean metric or imaginary
in Minkovski metric, the paper is using the Euclidean metric for convention. 

If the system is subject to parity violation, which is still an open
question in heavy-ion collision experiments, there are in general
other types of responses in the system besides Eq.(\ref{eq:current-correlation})
and Eq.(\ref{eq:spin-correlation}). For example in an effective theory
\cite{Damgaard:1993cr,Damgaard:1993sx} there may be response to a
pseudoscalar $\Phi^{ex}$ source coupled to quarks as $\bar{\psi}\gamma_{5}\Phi^{ex}\psi$,
leading to chirality imbalance. In such case, the effective fields
theory QGP is considered as a chiral superfluid discussed in Ref.\cite{Kalaydzhyan:2012ut,Kalaydzhyan:2014yda},
which is formally similar with a theory of topological superconductor
in condensed matter physics \cite{Qi:2012cs}. However, in the present
paper, we focus on an effective fields theory of QGP being parity
conserving, which is analogous to be another topological state of
matter, a topological insulator. The topological superconductor and
topological insulator are both fully gapped in the bulk for the quasi-particle
excitations, they are both topologically ordered state of matter in
the sense that their characters are topologically protected. A topological
superconductor and topological insulator actually can be turned into
each other by a phase transition discussed in Ref.\cite{Diamantini:2011gi}.
Based on the analogy, a parity conserving QGP in our paper is comparable
to the superfluidity model of QGP. There are essential differences
between them: (1) $\theta$ in the topological $\theta$-term is a
dynamical field in the effective chiral superfluid model which leads
to the chirality imbalance, but in our effective theory $\theta=\pi$
is a constant topological order parameter which does not violate parity.
(2) In a topological superconductor the gauge field and corresponding
fluid is massive while in a topological insulator the gauge field
and corresponding fluid remains gapless in the long wavelength limit.
(3) We will see in the next section that in our model the nearly perfect
transport properties is due to the self-duality of the fluids configurations
or BPS solution of the theory, but rather due to the superfluidity
of the fluids, since there is no symmetry breaking in our theory. 

It can be regarded that the Faddeev-Popov ghosts does not play fundamental
role in our effective theory in the sense that the building blocks
of the effective theory are the gauge invariant linear responses.
The system is written down to the lowest order which is linearized
and Abelianized, although the system is essentially non-linear. Surely
the non-linear interactions coming from the higher order terms may
be important to the properties of the system. The non-Abelian gluonic
fields in QCD defined at high energy makes no sense at the scale of
our interest and are effectively replaced by the linearized responses
or fluids, in this sense they can be well gauge fixed without Faddeev-Popov
ghosts. 

One may wonder that since the system is gapped, there seems no Fermi
surface. But remind that the system is bosonized, the chemical potential
$\mu$ can also be well-defined which formally shifts the non-dynamical
time component of gauge field $A_{0}^{ex}$, as if the quarks fill
the Fermi sphere in the equivalent bosonized QCD.

\section{General Features}

\subsection{Plasma with both Electric and Magnetic Charges}

At criticality, the fixed point coupling constant $g$ is large, then
the first two topological terms in Eq.(\ref{eq:effective-action})
dominate the low energy theory at the criticality, while the dynamical
Maxwell term can be seen as a perturbation. The classical equations
of motion of the topological action are given by,
\begin{align}
\frac{\delta S_{eff}}{\delta a_{\mu}}=0:\quad j_{\mu}-\frac{\theta}{2\pi}j_{\mu}^{m} & =0.\label{eq:EOM-1}
\end{align}
\begin{equation}
\frac{\delta S_{eff}}{\delta b_{\mu\nu}}=0:\quad f_{\alpha\beta}(a)-f_{\alpha\beta}(A^{ex})=0.\label{eq:EOM-2}
\end{equation}
The first equation of motion gives the electric current in terms of
the magnetic monopole current $j_{\mu}^{m}=\frac{1}{2\pi}\partial_{\nu}\tilde{f}_{\mu\nu}$
in the fluid \cite{witten1979dyons}, the monopole taking magnetic
charge receives an electric charge $q_{e}=q\theta/2\pi$, where $q$
is the electric charge of the system conventionally defined in the
covariant derivative. In the theory with $\theta=\pi\neq0$, the fluid
carries both electric and magnetic charges \cite{liao2007strongly,chernodub2007magnetic}.
As is well known that in the 2+1 dimensional topological Chern-Simons
theory gauge field plays the role of attaching fluxes to electrons,
in 3+1 dimensions the gauge field $a_{\mu}$ is attaching a monopole
to the electron (known as dyon). We will see in the latter discussion
that this dyonic property is crucial for a small value of shear viscosity
over entropy density of the QGP. The second equation of motion is
a constraint because the $b$ field plays the role of a Lagrangian
multiplier, or equivalently, the mass of the $b$ field can be considered
infinitely heavy shown as Eq.(\ref{eq:kinetic-b}). The second equation
of motion relates the configurations $a$ and $A^{ex}$.

\subsection{Interpretation of the theory via Hydrodynamics}

In this subsection, we discuss the physical interpretation of the
effective gauge fields $a_{\mu}$ and $b_{\mu\nu}$. Unlike the 1-form
gluonic fields defined at asymptotically free regime, in which a weakly
coupled photon analogous interpretation can be directly borrowed.
However, what is the gauge fields $a_{\mu}$, and even the 2-form
$b_{\mu\nu}$ in the effective theory? 

As is discussed in Section II, the topological dominant effective
field theory describes a hydrodynamic theory for two types of topological
fluids, the charge current and spin current, without mentioning their
microscopic origins. Here we have 1+6 continuity equations, one continuity
equation for the charge current Eq.(\ref{eq:charge-current-conservation})
and 6 continuity equations for the spin currents Eq.(\ref{eq:spin-current-conservation}),
\begin{equation}
\partial_{\mu}j_{\mu}=0,\quad\partial_{\alpha}M_{[\mu\nu]\alpha}=0.
\end{equation}
 In general, there is a relation connecting an energy-momentum tensor
to the spin current 
\begin{equation}
M_{[\mu\nu]\alpha}(x)=l_{\mu}T_{\nu\alpha}(x)-l_{\nu}T_{\mu\alpha}(x),
\end{equation}
in which $l$ is a characteristic size of the spin vortex. By using
the relation, we could construct a energy-momentum tensor via the
effective spin current without concerning the gluon degrees of freedom
making sense at high energy. The continuity equations for the spin
current are given by
\begin{equation}
\partial_{\alpha}M_{[\mu\nu]\alpha}=T_{\nu\mu}+l_{\mu}\partial_{\alpha}T_{\nu\alpha}-T_{\mu\nu}-l_{\nu}\partial_{\alpha}T_{\mu\alpha}=0.
\end{equation}
Under the condition that the energy-momentum tensor is symmetric $T_{\mu\nu}=T_{\nu\mu}$,
one can recognize that the 6 continuity equations for the spin currents
naturally contain 4 continuity equations for energy-momentum, 
\begin{equation}
\partial_{\mu}T_{\mu\nu}=0.\label{eq:stress-conservation}
\end{equation}

In addition, the classical equation of motion Eq.(\ref{eq:EOM-1})
$j_{\mu}=\frac{\theta}{2\pi}M_{[\mu\nu]\nu}$ implies a relation between
the current and the energy-momentum tensor so defined, for $\theta=\pi$,
we have $T_{\mu\nu}=\partial_{\mu}j_{\nu}+\partial_{\nu}j_{\mu}+g_{\mu\nu}T_{\alpha\alpha}$.
The relation naturally leads to a traceless condition $T_{\alpha\alpha}=0$,
so the stress-energy tensor is related to the current by 
\begin{equation}
T_{\mu\nu}=\partial_{\mu}j_{\nu}+\partial_{\nu}j_{\mu}.\label{eq:T-j}
\end{equation}
For the zero and near-zero modes of the fluids well below the gap,
we can define a potential (curl-free) flow vector $v_{\mu}$ by the
charge current as 
\begin{equation}
j_{i}=\rho v_{i}=-D\partial_{i}\rho,\label{eq:definition-j}
\end{equation}
where $D$ is a diffusion constant with dimension of length and $\rho=j_{0}$
is the flow density, we obtain 
\begin{equation}
\partial_{0}\left(\rho v_{j}\right)=D\partial_{i}T_{ij},\quad(i,j=1,2,3).\label{eq:v-DT}
\end{equation}
By using Eq.(\ref{eq:T-j} and \ref{eq:definition-j}), we have 
\begin{equation}
D\partial_{i}T_{ij}=D\partial_{i}\left[\partial_{j}\left(\rho v_{i}\right)+\partial_{i}\left(\rho v_{j}\right)\right]=-\partial_{i}\left(\rho v_{i}v_{j}\right)+D\rho\nabla^{2}v_{j},
\end{equation}
so the Eq.(\ref{eq:v-DT}) (the classical equation of motion Eq.(\ref{eq:EOM-1}))
are nothing but surprisingly the Navier-Stokes equations for incompressible
fluid without pressure,
\begin{equation}
\partial_{0}\left(\rho v_{j}\right)+\partial_{i}\left(\rho v_{i}v_{j}\right)=\eta\nabla^{2}v_{j},\quad(i,j=1,2,3).\label{eq:NS-equ}
\end{equation}
in which $\eta=D\rho$ here is a viscosity constant.

In summary, we show the close relation between the effective field
theory and hydrodynamic equations. The ground state of the theory
can be interpreted as a theory of topological fluids satisfying hydrodynamics
equations, including the continuity equation of current, the conservation
of the energy-momentum tensor Eq.(\ref{eq:stress-conservation}),
the traceless of the energy-momentum tensor, and the Navier-Stokes
equations Eq.(\ref{eq:NS-equ}). The surprising relations shed some
light on the fact (i) mentioned in the introduction, namely, why the
long distance behavior of QGP around the critical temperature is so
well described by hydrodynamics.

\subsection{Nearly Perfect Liquid}

It is worth stressing that since the low energy effective theory is
dominated by the topological terms, so in principle, there is no strict
concepts of metric and energy, one may wonder why we can talk about
the the energy-momentum tensor defined above. The reason is that the
energy-momentum tensor defined above just describes classical flows
or distributions of the charge and spin currents in the ground state
gapped from the excitations, which are perfect fluids. Like the Hall
fluids which have vanishing viscosity, if the low energy behavior
of quark-gluon fluids here are solely governed by the topological
terms, they have no viscosity either (completely perfect). It is a
general property that the dissipationless of the topological term
does not depend on that it is in Euclidean or Minkovski formalism.
When the excitation states coming from the non-topological terms,
i.e. Maxwell term, are taken into account, internal frictions appear
and the fluid becomes dissipation. Strictly speaking, dissipation
or broken time reversal appears only in Masubara but Minkovski formalism
of the non-topological part of the theory. At temperature well below
the gap which protects the topological order, the deviation from viscousless
is small, so the fluids still behave nearly perfect. This property
of the theory is associated with the fact (ii).

The terminology ``incompressible'' of the fluids is tantamount to
``topological'', more precisely, since the partition function given
by the topological fixed point action only depends on the topology
of the manifold, so it is exactly responseless to the compressing
or expanding of the 3-volume of the manifold. As a result, the quantities
like pressure and bulk viscosity of QGP are identically zero predicted
from the lowest order topological theory.

To consider the transport beyond the lowest order result, one can
first integrate out the $a$ and $b$ field, the effective strongly
coupled fixed point action becomes
\begin{equation}
S_{eff}(A^{ex})=\frac{i\theta}{8\pi^{2}}\int d^{4}x\epsilon_{\mu\nu\alpha\beta}\partial_{\mu}A_{\nu}^{ex}(t,\mathbf{x})\partial_{\alpha}A_{\beta}^{ex}(t,\mathbf{x})+\frac{1}{4g^{2}}\int d^{4}xF_{\mu\nu}(A^{ex})F^{\mu\nu}(A^{ex})+\mathcal{O}\left(\partial^{4}/\Lambda^{2}\right),\label{eq:action-by-external-A}
\end{equation}
which at lowest order can be viewed as an action for the U(1) electromagnetic
fields in a QGP medium, complicated non-linear self-interactions are
in the higher order terms. 

The $\theta$ parameter in the first $\theta$-term is inherited from
its ancestor action Eq.(\ref{eq:effective-action}) which is non-trivial
$\theta=\pi$. It may come as a surprise that there is an extra topological
$\theta$-term at leading order in the effective U(1) electromagnetic
fields in the QGP medium, the reason we will see is closely related
to the dyonic nature of the $a$ fluids of the QGP medium. Because
at the classical level the constraint Eq.(\ref{eq:EOM-2}) suggests
a close relationship between the $A$ fields and the dyonic $a$ fields,
non-trivial topology of the $A$ fields is also expected which makes
the non-trivial $\theta$-term should not be simply removed as an
ordinary QED. Such result can be understood intuitively as that by
intermediating the dyonic medium the electric fields now are able
to interact with magnetic fields effectively (magnetoelectric effect). 

The second term is the dynamical Maxwell term which is also expected
being suppressed at low energy since the integration over QCD fields
gives a Debye gap. The transport of the effective theory e.g. the
conductivity, charge susceptibility and shear viscosity, therefore
take small values that are of order of the dynamical Maxwell term.
The following calculations are doing at zero temperature, and the
finite temperature results can be generalized by standard Matsubara
summation procedures.

Since the fluid is incompressible if we only consider the topological
term, the topological term only gives a current transport on the boundary
of the QGP, there is no net current in its bulk. The current transport
or conductivity in the bulk solely gives rise from the Maxwell term,
\begin{equation}
\sigma=\lim_{\omega\rightarrow0}\frac{1}{\omega}\frac{\delta^{2}S_{eff}(A^{ex})}{\delta A_{i}^{ex}(\omega,k=0)\delta A_{j}^{ex}(0,0)}P_{ij}^{\bot}=\lim_{\omega\rightarrow0}\frac{1}{2g^{2}}\frac{\omega^{2}}{\omega}=\frac{\omega}{2g^{2}},
\end{equation}
in which $P_{ij}^{\bot}=\delta_{ij}-k_{i}k_{j}/\left|\mathbf{k}\right|^{2}$
is the transverse projector. Note that the chemical potential $\mu$
can be seen as a shift of the non-dynamical $A_{0}^{ex}$, i.e. $A_{0}^{ex}(\omega=0,\mathbf{k})$,
then the charge susceptibility can be calculated by
\begin{equation}
\chi=\frac{\delta^{2}S_{eff}}{\delta\mu^{2}}=\lim_{\left|\mathbf{k}\right|\rightarrow0}\frac{\delta^{2}S_{eff}(A^{ex})}{\delta A_{0}^{ex}(\omega=0,\mathbf{k})\delta A_{0}^{ex}(0,0)}=\frac{\left|\mathbf{k}\right|^{2}}{2g^{2}}.
\end{equation}
One can find that the ratio between the conductivity and susceptibility
matches the Einstein relation,
\begin{equation}
\frac{\sigma}{\chi}=\frac{\omega}{\left|\mathbf{k}\right|^{2}}=D,\label{eq:einstein-relation}
\end{equation}
in which $D$ is the charge diffusion constant. For a special interest,
the value of $D$ can be determined on self-dual configurations shown
as follows.

An important observation to the effective theory is that a lower bound
can be realized from the non-trivial self-dual configuration $F_{\mu\nu}=\tilde{F}_{\mu\nu}=\frac{1}{2}\epsilon_{\mu\nu\rho\sigma}F_{\rho\sigma}$.
It is known that there is a Bogomol'nyi bound at the self-dual configuration,
\begin{equation}
\frac{1}{4g^{2}}\int d^{4}xF_{\mu\nu}F^{\mu\nu}\geqslant S_{S.D.}\overset{\mathrm{Self-Dual}}{\equiv}\frac{8\pi^{2}\left|Q\right|}{g^{2}},
\end{equation}
where $Q=\frac{1}{16\pi^{2}}\int d^{4}xF_{\mu\nu}\tilde{F}_{\mu\nu}=\frac{1}{16\pi^{2}}\int d^{4}xf_{\mu\nu}(a)\tilde{f}_{\mu\nu}(a)\in\mathbb{Z}$
and in which the second equal sign is given by the constraint Eq.(\ref{eq:EOM-2}).
So the result could be understood from the fact that the self-duality
of $A^{ex}$ fields is closely connected to the self-duality of the
dyonic fluids $a_{\mu}$ fields via the constraint, the integer is
effectively relates to the topological charge of the $a_{\mu}$ fluids
configurations.

As a first consequence of the self-duality, it leads to a relation
$\omega^{2}=\left|\mathbf{k}\right|^{2}$ in Euclidean metric for
the self-dual zero mode. Thus the charge diffusion constant $D$ in
Eq.(\ref{eq:einstein-relation}) equals to $D=1/\omega$ on the self-dual
configurations, which at the lowest Matsubara frequency $\omega=2\pi T$
takes value $D=\frac{1}{2\pi T}$, agreeing with \cite{policastro2002ads}.
The smallness of the diffusion constant is because it stems from the
zero mode of the fluids, but rather the excited states.

If we set $g=4\pi/N_{c}$ and replace $\omega$ by the lowest Matsubara
frequency $\omega=2\pi T$, then the conductivity and susceptibility
agree with \cite{policastro2002ads}
\begin{equation}
\sigma=\frac{N_{c}^{2}T}{16\pi},\quad\chi=\frac{N_{c}^{2}\left|\mathbf{k}\right|^{2}T}{16\pi\omega}.
\end{equation}

The second consequence of the self-duality is about the shear viscosity,
which is related to component of correlation function of stress-energy
tensor being a response to metric. Differ from the topological terms,
the Maxwell term is the only term in the effective action containing
metric $g_{\mu\nu}$. Strictly speaking, all physics of QCD are encoded
in $Z[a]$, to derive shear viscosity of quark-gluon liquids, we need
to vary with respect to the metric perturbations of Maxwell term in
$Z[a]$, but when $a$ and $b$ fields are integrated out, it is equivalent
to vary with respect to the metric perturbation in the effective action
$S_{eff}(A^{ex})=-\ln Z[A^{ex}]$. We will rewrite the action by explicitly
including the metric (in order to distinguish between the gauge coupling
and metric, we use $g_{*}$ to represent the gauge coupling),
\begin{equation}
S=\frac{1}{4g_{*}^{2}}\int d^{4}x\sqrt{g}F_{\alpha\beta}F^{\alpha\beta},
\end{equation}
then doing the functional derivative with respect to the metric and
finally taking the limit of flat metric back 
\begin{equation}
\lim_{\sqrt{g}\rightarrow1}\frac{\delta^{2}S_{eff}(A^{ex})}{\delta g_{xy}\delta g_{xy}}=\frac{1}{4g_{*}^{2}}\left(F_{xy}F_{xy}-\frac{1}{2}F_{\alpha\beta}F^{\alpha\beta}\right).
\end{equation}

By considering the QGP is homogeneous and isotropic ($\left|F_{xy}\right|=\left|F_{xz}\right|=\left|F_{yz}\right|$),
so that the component satisfies $F_{xy}F_{xy}=\frac{1}{6}F_{\alpha\beta}F^{\alpha\beta}$.
For the field strengths in the action are homogeneous, the two-point
correlation function of stress-energy tensor $G_{xy,xy}$, is just
given by the average action density,
\begin{equation}
G_{xy,xy}(\omega,k)=\lim_{\sqrt{g}\rightarrow1}\frac{\delta^{2}\ln Z[A^{ex}]}{\delta g_{xy}(\omega,k)\delta g_{xy}(0,0)}\geqslant\frac{8\pi^{2}\left|Q\right|}{3g_{*}^{2}V_{4}},
\end{equation}
in which $V_{4}$ is a Euclidean 4-volume defined as $\ln Z(x)=V_{4}\int\frac{d^{4}k}{(2\pi)^{4}}\ln Z(k)$.
Then, by using the Kubo formula, the minimal value of action gives
lower bound to a shear viscosity
\begin{align}
\eta & =\lim_{\omega\rightarrow0}\frac{1}{2\omega}G_{xy,xy}(\omega,k)\geqslant\frac{2\pi\left|Q\right|}{3g_{*}^{2}V_{3}}.
\end{align}
in which we have used $\lim_{\omega\rightarrow0}\frac{1}{\omega V_{4}}=\frac{1}{2\pi V_{3}}$.

On the other hand, the third law of thermodynamics tells us that there
are non-vanishing entropy near zero temperature. Considering the thermodynamic
relation for the entropy density $s=\partial p/\partial T$, moreover,
because the the system is isotropic, $p_{x}=p_{y}=p_{z}=p$, then
the pressure is related to the energy density by $\epsilon=3p$, which
can be given by the minimal value of the action $\epsilon=-\frac{T}{V_{3}}\ln Z=\frac{T}{V_{3}}S_{S.D.}$
near the zero temperature. Finally we have an asymptotic entropy density
in the vicinity of the zero temperature 
\begin{equation}
s=\frac{8\pi^{2}\left|Q\right|}{3g_{*}^{2}V_{3}}.
\end{equation}
The entropy density averaged shear viscosity is a dimensionless quantity,
which measures a pure quantum (zero temperature) originated intrinsic
viscosity,
\begin{equation}
\frac{\eta}{s}=\frac{1}{4\pi},
\end{equation}
in which is archived when the system is self-dual, or equivalently
in a BPS state. 

An important observation of the section is that the self-duality here
is crucial for not only achieving the small value of diffusion constant
$D$ but also for that of the $\eta/s$. It reflects the fact that
the small value of these quantities are closely related to a mixture
of equal importance of the electric and magnetic components of the
plasma, which has been discussed in \cite{liao2007strongly,liao2008magnetic-1,liao2008magnetic-2,liao2008magnetic-3}.
This small value of $\eta/s$ is also obtained from another self-dual
theory \cite{policastro2001shear,policastro2002ads}, the AdS/CFT
correspondence, in which people conjectures that the constant is a
universal lower bound. The assumption of isotropy is also important
for the result, since without isotropic the lower bound of the action
can not constrain each component of the shear viscosity, and hence
certain component may be lower than $1/4\pi$ and breaks the conjecture.

\subsection{Fractionalized Charge}

In this section we show that the fact (iii) is also a natural consequence
of the effective theory. Charge fractionalization is an important
mathematical feature of the topological action, and hence be a characteristic
of the QGP phase being a topological phase. Following the form of
the effective action Eq.(\ref{eq:effective-action}), we drop the
Maxwell term and focus on the topological action. Considering effective
field action for each parton field (labelled by flavor index $I=1,2,...,\mathsf{m}$)
are in its fractionalized topological order characterized by coefficient
$\mathsf{m}$,
\begin{align}
S^{(I)} & =\frac{i\mathsf{m}}{2\pi}\int d^{4}x\epsilon_{\mu\nu\alpha\beta}b_{\mu\nu}^{(I)}\partial_{\alpha}a_{\beta}^{(I)}+\frac{i\theta}{8\pi^{2}}\int d^{4}x\epsilon_{\mu\nu\alpha\beta}\partial_{\mu}a_{\nu}^{(I)}\partial_{\alpha}a_{\beta}^{(I)}+\int d^{4}xj_{\mu}^{(I)}A_{\mu}^{ex}
\end{align}
in which $j_{\mu}^{(I)}=\frac{\mathsf{m}}{2\pi}\epsilon_{\mu\nu\alpha\beta}\partial_{\nu}b_{\alpha\beta}^{(I)}$
is the parton current. Like the Chern-Simons theory of the fractional
quantum Hall fluids in which the Chern-number characterizes its topological
order, the fractionalized topological order of each parton in the
plasma is characterized by the integer number $\mathsf{m}$ in its
action, which must be odd for the fermion statistic of each parton. 

Since the effective coefficient $\mathsf{m}_{eff}$ in an effective
action for total partons follows the composition law $\mathsf{m}_{eff}^{-1}=\sum_{I}\mathsf{m}_{I}^{-1}$
from the coefficient of each parton, hence we can prove that the effective
action is given by
\begin{equation}
Z[A^{ex}]=\int\prod_{I}\mathcal{D}a^{(I)}\mathcal{D}b^{(I)}\exp\left(-\sum_{I}S^{(I)}\right)=\int\mathcal{D}a\mathcal{D}b\exp\left(-S_{eff}\right)
\end{equation}
in which
\begin{equation}
S_{eff}=\frac{i}{2\pi}\int d^{4}x\epsilon_{\mu\nu\alpha\beta}b_{\mu\nu}\partial_{\alpha}a_{\beta}+\frac{i\theta}{8\pi^{2}}\int d^{4}x\epsilon_{\mu\nu\alpha\beta}\partial_{\mu}a_{\nu}\partial_{\alpha}a_{\beta}+\int d^{4}xj_{\mu}A_{\mu}^{ex},
\end{equation}
with fractionalized charges and parton fields satisfying 
\begin{equation}
e=\sum_{I}e_{I}=1,\quad a_{\mu}=\sum_{I}a_{\mu}^{(I)},\quad b_{\mu\nu}=\sum_{I}b_{\mu\nu}^{(I)},\quad j_{\mu}=\sum_{I}j_{\mu}^{(I)}.
\end{equation}

The observed effective charge $e$ is quantized by fundamental unit,
but here the current $j_{\mu}$ seems as a composite current and can
be splited into constitutes with fractionalized charges $e_{I}=e/\mathsf{m}$
in its fractionalized topological order. In this sense, the plasma
phase is a deconfined phase. In the soup of QGP, the baryon with integer
electric charge is splited into three individual quarks with fractionalized
electric charges. In this framework, the exotic assumption that quarks
have fractional electric charge ($e/3$) is a natural consequence
of the deconfined topologically ordered phase characterized by the
odd integer $\mathsf{m}=3$.

Reminding that there are two independent fluids in the plasma, the
charge current and spin current. It is worth noting that the fractionalization
does not apply to the spin current. Opposite to the direct coupling
between the charge current and the external gauge field $A^{ex}$,
the coupling between the spin density and the external magnetic field
is indirect, mediating by the infinitely massive $b$ fields. The
$b$ field, as is proved, could be splited into constitutes $b^{(I)}$
and hence the corresponding coupling between $b^{(I)}$ and external
magnetic field $\tilde{F}^{ex}$ is fractionalized, however, the $\tilde{F}^{ex}$
can not couple to each individual spin constitute $s_{\mu\nu}^{(I)}=\frac{\mathsf{m}}{2\pi}\epsilon_{\mu\nu\alpha\beta}\partial_{\alpha}a_{\beta}^{(I)}$
since all intermediate constitute $b^{(I)}$ fields must be integrated
out. Thus the coupling constant between spin density and external
magnetic field is not fractionalized into smaller units, i.e. spin
or angular momentum does not split.

\section{Conclusions}

In this paper, we show remarkable connections between a low energy
effective field theory of QGP and a topological quantum field theory,
and a hydrodynamic theory. In many aspects the deconfined QGP phase
of strong interaction system is analogous to the topological phase
in condensed matter systems. In fact, the central argument of the
paper is that it is indeed a topological phase which is also argued
in \cite{Zhitnitsky:2013hs}. The low energy degrees of freedoms in
the topological phase are incompressible quantum fluids which are
analogous with the nearly perfect quark-gluon liquid around $T_{c}$
discovered at RHIC. Bosonization approach makes an effective description
to the topological phase and low energy topological fluid possible,
due to the fact that a gauge theory with massive fermions and the
bosonized description share the same topological limit. Below the
massive scale, which is expected several times of $T_{c}$, the bosonization
approach is valid, and the effective action Eq.(\ref{eq:effective-action})
is the central result of the paper, which is suggested as an infrared
fixed point action for the QCD system coupling to an U(1) external
source. A hydrodynamic interpretation of the gauge theory requires
a non-trivial value $\theta=\pi$. In such a theory, the QGP is effectively
described by two types of fluids: charge current and spin currents,
governed by hydrodynamics continuity equations. They represent the
low energy collective modes about the topologically ordered state.
The exotic phenomenon of QGP discovered at RHIC, for instance the
smallness of pressure and viscosity around $T_{c}$, as well as the
charge fractionalization of the excitations, seem much easier to be
understood by its topological nature suggested in the effective theory.
Both the electric and magnetic components are equally important and
half-half mixing in the quark-gluon liquids, which is also topologically
originated. Indeed, the electric-magnetic self-duality is closely
related to achieving a small value of diffusion constant and $\eta/s$,
making the quark-gluon liquids most perfect liquids.
\begin{acknowledgments}
This work was supported in part by the National Science Foundation
of China (NSFC) under Grant No.11205149, and Science Research Foundation
of Jiangsu University under Grant No.15JDG153.
\end{acknowledgments}

\bibliographystyle{apsrev}

\end{document}